# Negative-charge-storing mechanism of potassium-ion electrets used for vibration-powered generators: Microscopic study of a-SiO$_2$ with and without potassium atoms

Toru Nakanishi,[1,a)] Takeshi Miyajima,[1] Kenta Chokawa,[2,b)] Masaaki Araidai,[2,1] Hiroshi Toshiyoshi,[3] Tatsuhiko Sugiyama,[4] Gen Hashiguchi,[4] and Kenji Shiraishi[2,1]

[1]*Graduate School of Engineering, Nagoya University, Nagoya 464-8603, Japan*
[2]*Institute of Materials and Systems for Sustainability, Nagoya University, Nagoya 464-8601, Japan*
[3]*Institute of Industrial Science, The University of Tokyo, 4-6-1 Komaba, Meguro-ku, Tokyo 153-8505, Japan*
[4]*Research Institute of Electronics, Shizuoka University, 3-5-1 Johoku, Naka-ku, Hamamatsu, Shizuoka 432-8011, Japan*
(Dated: 09 September 2020)

A potassium-ion electret, which is a key element of vibration-powered microelectromechanical generators, can store negative charge almost permanently. However, the mechanism by which this negative charge is stored is still unclear. We theoretically study the atomic and electronic structures of amorphous silica (a-SiO$_2$) with and without potassium atoms using first-principles molecular-dynamics calculations. Our calculations show that a fivefold-coordinated Si atom with five Si–O bonds (an SiO$_5$ structure) is the characteristic local structure of a-SiO$_2$ with potassium atoms, which becomes negatively charged and remains so even after removal of the potassium atoms. These results indicate that this SiO$_5$ structure is the physical origin of the robust negative charge observed in potassium-ion electrets. We also find that the SiO$_5$ structure has a Raman peak at 1000 cm$^{-1}$.

Energy-harvesting technology, which converts energy in the environment from sources such as light, heat, and vibration into electrical power, is a key element in meeting energy demands.[1–4] Many types of energy-harvesting devices have been proposed, and these include thermoelectric devices and vibration-powered generators. Such devices tend to have maximum power outputs of several μW to a few mW, and they therefore have had limited conventional uses. However, in recent years, the range of applications for energy-harvesting devices has expanded due to the development of low-power technology. In particular, maintenance-free autonomous power supplies are very important for the Internet of Things and the Trillion Sensors initiative. Energy-harvesting technology has therefore attracted a significant amount of attention in relation to these technologies.

Among the many energy-harvesting technologies, vibration-powered generation is expected to provide a stable supply of electricity because its output does not depend on specific energy sources in the natural environment such as sunlight. Accordingly, vibration-powered generators can be widely used as stand-alone power sources for sensors in environments with vibration, such as transportation machinery, roads, and wearable devices.[5] Recently, a vibration-powered microelectromechanical generator with a potassium-ion electret has attracted increasing attention.[6–11] A potassium-ion electret can be fabricated by adding potassium atoms to amorphous silica (a-SiO$_2$) and then subsequently removing them. A potassium-ion electret can permanently store a negative charge, and this enables the creation of maintenance-free power-generation devices. However, the mechanism by which this negative charge is stored in potassium-ion electrets is still unclear. In general, the Si atoms in a-SiO$_2$ form four Si–O bonds, resulting in a fourfold coordinated structure having no characteristic defects with negative charges. Hence, it is important to clarify the role of potassium atoms in forming negatively charged defects in a-SiO$_2$.

In this study, we simulated the manufacturing processes of potassium-ion electrets using first-principles calculations and created a-SiO$_2$ structures containing potassium atoms. We were then able to reveal the characteristic local structures formed by the presence of potassium atoms, and we examined the charge state in a local structure to clarify the origin of the negative charges. We were able to confirm that the local structures, which are the origin of the negative charges, are retained even after the removal of the potassium atoms. Finally, we propose guidelines for identifying the characteristic local structure.

The calculations were performed using the Vienna Ab initio Simulation Package (VASP), which is based on density functional theory.[12–15] In the structural optimization, we stopped the calculations when the maximum force on each atom was less than $5.0 \times 10^{-2}$ eV/Å. The nuclei and core electrons were simulated by pseudopotentials generated by the projector augmented-wave method.[16] The valence wave functions were expanded using the plane-wave basis set, for which we found that a cut-off energy of 460 eV suffices. Twelve **k** points in a $2 \times 2 \times 3$ Monkhorst–Pack grid were sampled for the Brillouin zone integrations. We used a GGA-PBE-type exchange–correlation functional[17] for the molecular dynamics (MD) calculations and structural optimization. The Heyd, Scuseria, and Ernzerhof (HSE) exchange–correlation functional[18] was also used to obtain an accurate value for the energy of each structure. We set the fraction of the Fock exchange and the range-separation parameter in the HSE functional to be 0.51 and 0.20 Å$^{-1}$, respectively, which reproduces the bandgap of 9.0 eV for SiO$_2$. The volume of a supercell was fixed during all the calculations. We used a Bader charge analysis[19–22] to investigate the charge states.

a) Electronic mail: nakanishi.toru@e.mbox.nagoya-u.ac.jp
b) Author to whom correspondence should be addressed:
chokawa@fluid.cse.nagoya-u.ac.jp

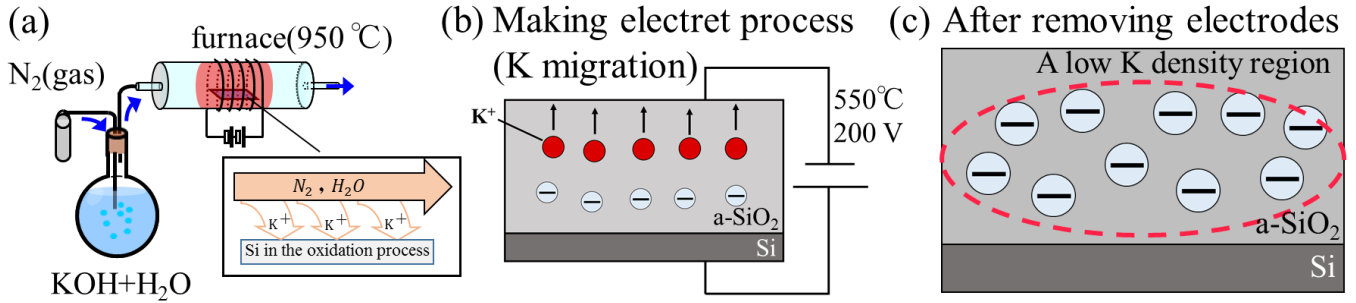

**FIG. 1.** Schematic illustration of fabrication of K-ion electrets.[6,24-25] (a) Thermal oxidation, which incorporates K atoms into a-$SiO_2$. (b) Charge-storage mechanism of a K-ion electret. On applying a voltage while heating, the K ions move to the cathode. (c) The cathodes are removed. The low-K-density region is negatively charged.

The VESTA system was used to draw the atomic configurations.[23]

According to experimental reports, potassium-ion electrets are made using the steps illustrated in Fig. 1.[6,24–25] First, a Si substrate is thermally oxidized by steam supplied through a KOH solution to obtain a-$SiO_2$ with potassium atoms. In this process, potassium and hydrogen atoms are incorporated into the a-$SiO_2$. Thereafter, a high voltage (200 V) is applied while heating the a-$SiO_2$ with potassium atoms at 550°C. During this process, the potassium ions move toward the cathode, and the low-potassium-density region becomes negatively charged. Finally, a negatively charged electret is fabricated by removing the cathode.

We created an a-$SiO_2$ structure with potassium atoms using the following steps. We prepared supercells of alpha quartz (144 atoms) and inserted potassium and hydrogen atoms into each supercell. The insertion positions were randomly chosen and resulted in four initial structures. The density of a-$SiO_2$ was set to the experimentally obtained value of 2.2 g/cm$^3$.[26] To obtain the amorphous structures, we performed melt–quench simulations using first-principles MD calculations. The temperature history used in these MD calculations was as follows. During the first 10 ps, the supercell was held at 5000 K to melt the structure, and it was then quenched to 0 K at a rate of 50 K/ps. The melt–quench technique has been extensively used in previous research to investigate a-$SiO_2$.[27–31] It has been reported that a-$SiO_2$ can be formed at a cooling rate of 200 K/ps, and almost the same structures are obtained with a cooling rate of 100 K/ps.[32] Thus, the cooling rate we used is sufficiently slow to form a-$SiO_2$. After the MD calculations, we performed structural-optimization calculations for each structure obtained and examined the resulting characteristic structures. Thereafter, we removed the potassium atoms from the supercells and re-performed the MD calculations to anneal these models. The temperature history used in this second set of MD calculations was as follows. During the first 10 ps, the supercell was held at 1000 K, and it was then quenched to 0 K at a rate of 50 K/ps. After this, we again optimized each charged structure.

We begin with examinations of the structures obtained after the first MD calculations. The structure of a-$SiO_2$ with a potassium atom is shown in Fig. 2(a). We found that all four a-$SiO_2$/potassium-atom models had two characteristic structures: a fivefold-coordinated Si atom and a Si–Si bond structure. These special structures appear even though the potassium atom is not involved in their bonding. To investigate the effect of the potassium-atom insertion, we also created a simple a-$SiO_2$ structure using the same temperature history as the initial MD calculations. In this structure, all the Si atoms had four Si–O bonds and there was no characteristic local structure, as noted in previous reports.[29] Therefore, these local structures, a fivefold-coordinated Si atom and a Si–Si bond, are quite different from those of ordinary a-$SiO_2$ and are considered to be induced by the potassium atom.

Figure 2(b) shows the detailed structure around a potassium atom. The ionic valences obtained by a Bader charge analysis are also depicted. The interatomic distance between the potassium and oxygen atoms is about 3 Å, and the potassium and oxygen atoms do not form any bonds. From the Bader charge analysis, we found that the ionic valence of a potassium atom is +0.9 [Fig. 2(b)], indicating that it behaves as a positively charged ion. However, the ionic valence of the O atoms around the potassium atom is about −1.6, which is the same as the ionic valence in ordinary $SiO_2$. From the above discussion, it is considered that the potassium atom behaves as a mobile cation, and this result is in good agreement with experiments, in which the potassium atoms can move toward the cathode when a voltage is applied.[6]

Figure 2(c) shows the detailed structure around the fivefold-coordinated Si atom with five Si–O bonds. Previous studies have suggested that fivefold-coordinated Si atoms exist in a-Si[33,34] and H-incorporated a-$SiO_2$.[35] Since a Si atom usually has four bonds, the presence of a fivefold-coordinated Si atom with five Si–O bonds is quite characteristic. We call this structure an $SiO_5$ structure. The usual fourfold-coordinated structure is an $SiO_4$ structure. The ionic valence of a Si atom and five O atoms in an $SiO_5$ structure are about +3.2 and −1.6, which is almost the same as the values in an $SiO_4$ structure. Since each O atom forms two Si–O bonds, the ionic valence of an O atom per Si atom is −0.8. Accordingly, the total ionic valence of an $SiO_4$ structure is found to be 0.0, indicating that $SiO_4$ structures are electrically neutral. However, since an $SiO_5$ structure consists of a Si atom with an ionic valence of +3.2 and five oxygen atoms, each with an ionic valence of −0.8, the total ionic valence of an $SiO_5$ structure is −0.8. Thus, we find that the $SiO_5$ structure is negatively charged in a-$SiO_2$ with a potassium atom.

Figure 2(d) shows the detailed structure around the Si–Si bond structure. The ionic valences of Si and O atoms are +2.4

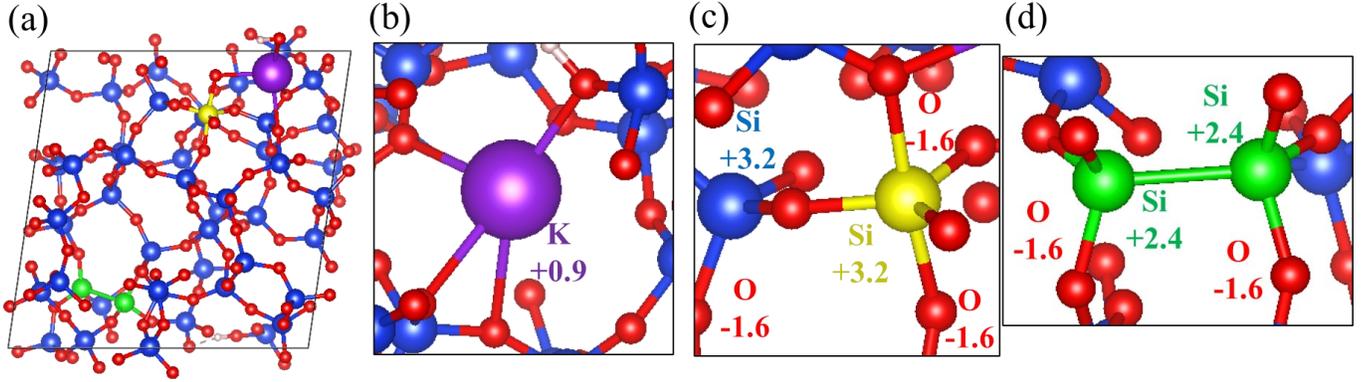

**FIG. 2.** (a) The a-SiO$_2$ structure obtained by the initial MD calculations. Some local structures are formed by inserting a K atom. Blue, red, and purple spheres denote Si, O, and K atoms, respectively. The characteristic local structures are shown in different colors. The yellow spheres are fivefold-coordinated Si atoms, which have five Si–O bonds. The green spheres are Si atoms with Si–Si bonds. (b) Structure around a potassium atom. (c) Structure around a fivefold-coordinated Si atom. (d) Structure around a Si–Si bond. The ionic valences obtained from a Bader charge analysis are also depicted.

and −1.6, respectively. This structure consists of two Si atoms with an ionic valence of +2.4 and six O atoms, each with an ionic valence of −0.8. The total ionic valence of these atoms is therefore 0.0. This indicates that this Si–Si bond structure is electrically neutral. As discussed above, charge transfer from the potassium atom to the SiO$_5$ structure occurs, resulting in the presence of K$^+$ ions and negatively charged SiO$_5$ structures.

We now discuss the atomic and electronic structures of a-SiO$_2$ after potassium-atom removal. Experimentally, it has been reported that potassium-ion electrets are negatively charged after removal of their potassium atoms. However, it is well known that a stable charged state sensitively depends on the position of the Fermi level. To consider the relative stability of the neutral state and the negatively and positively charged states, we calculated the formation energy of the charged system ($E_{form}$) using[36,37]

$$E_{form} = \left(E_{defect}(q) + E_{quartz}(0)\right) - \left(E_{defect}(0) + E_{quartz}(q)\right) + \frac{3q^2}{10\varepsilon r_0} + \begin{cases} qE_F & (q \geq 0), \\ q(E_F - E_{gap}) & (q < 0), \end{cases} \quad (1)$$

where $E_{defect}$ is the energy of a crystal with a defect, $E_{quartz}$ is the energy of quartz SiO$_2$, $E_{gap}$ is the bandgap, $\varepsilon$ is the dielectric constant, $r_0$ is the radius of a sphere with the same volume as a unit cell, $q$ is the charge state, and $E_F$ is the Fermi level of the system measured from the top of the valence band.

Figure 3 shows the formation energy of each charged state, and it can be seen that positively or negatively charged states are always more stable than the neutral state. Hence, we focus on the negatively and positively charged states. For the negatively charged state, the SiO$_5$ structure remains in a-SiO$_2$ even after removal of the potassium atom from the supercell. We performed a Bader charge analysis for these local structures and found that the SiO$_5$ structure is negatively charged and that the Si–Si bond structure is electrically neutral. For the positively charged state, the Si–Si bond is broken, and two Si dangling bonds appear, although the fivefold-coordinated SiO$_5$ structure is preserved. The formation of Si dangling bonds upon hole injection is consistent with a previous report.[38] From the Bader charge analysis, it was found that the fivefold-coordinated SiO$_5$ structure still retains an electron and remains negatively charged even after electron removal. Next, we looked at the Si dangling bonds in the positively charged state. From the Bader charge analysis, we found that the ionic valence of threefold-coordinated Si is +3.1, which is almost the same as that of the fourfold-coordinated SiO$_4$ structure. Because Si atoms with a dangling bond have three Si–O bonds, the Si-dangling-bond structure is positively charged. Accordingly, we can conclude that the system is positively charged due to the two positively charged Si dangling bonds.

More importantly, it was found that the intersection between the positively and negatively charged states is located at 3.7 eV. This result means that the stable charge state of a-SiO$_2$ completely reverses between positively and negatively charged states at a threshold of 3.7 eV. To obtain stable negatively charged potassium-ion electrets, the system Fermi level should be greater than 3.7 eV in the process conditions. In other words, lowering the Fermi level induces the appearance of positively charged Si-dangling-bond structures.

We now discuss the Fermi level condition during the experimental conditions. In the fabrication process of an electret,[6] there is a Si substrate below the potassium-ion electret itself [Fig. 1(b)]. Thus, it is estimated that the position of the Fermi level in the SiO$_2$ is determined by the Fermi level position of the Si substrate, and the band offset at the Si/SiO$_2$ interface has already been reported by a theoretical study (Fig. 4).[39] Since the Fermi level of Si is within the bandgap, the Fermi level at the Si/SiO$_2$ interface is located between 4.4 eV and 5.5 eV above the top of the valence band of SiO$_2$. In this region, the negatively charged state becomes the most stable configuration (Fig. 4). Consequently, experimentally obtained potassium-ion electrets with Si substrates spontaneously form SiO$_5$ structures and store a negative charge. It is expected that this negative charge can be almost permanently retained due to the high strength of Si–O bonds, and that potassium-ion electrets should therefore enable the realization of high-reliability vibration-powered generators. Moreover, the higher the Fermi level, the more stable the negatively charged state. Therefore, we can conclude that a potassium-ion electret produced from a n-type Si substrate will be more reliable than one produced from a p-type Si substrate. Moreover, using an n-type SiC substrate is also promising because its Fermi level is located higher than that of an n-type Si substrate.

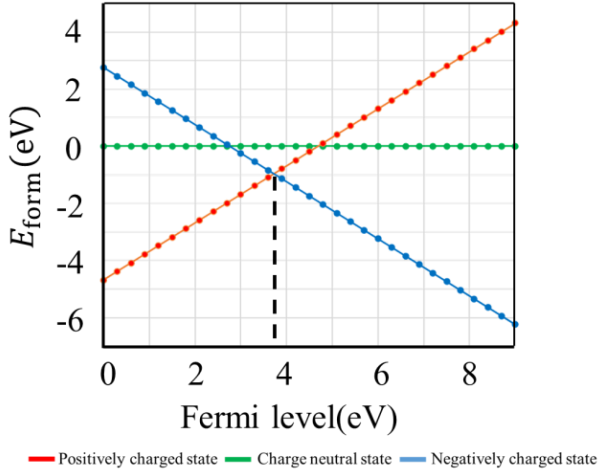

**FIG. 3.** Formation energy diagram after removal of the K$^+$ ion. The negatively charged state becomes stable when the Fermi level is higher than 3.7 eV. We set the energy of the neutral charged state to 0.

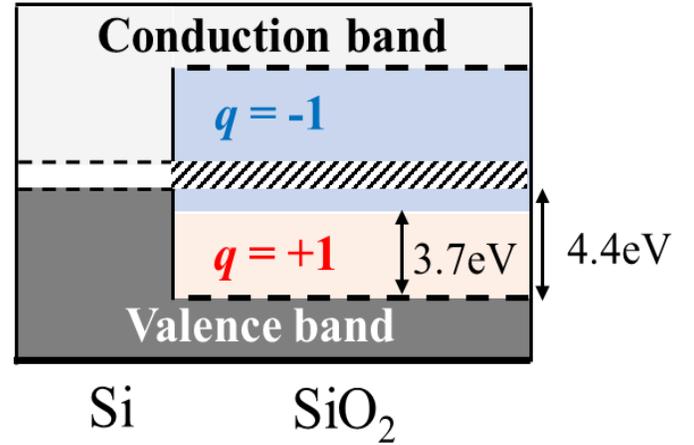

**FIG. 4.** Schematic band diagram of the Si/SiO$_2$ interface with the stable charged state. The Fermi level at the Si/SiO$_2$ interface is dominated by the Fermi level of Si. The striped area is the Fermi level of the system. The negatively charged state is always stable when we use Si substrates.

Finally, we consider guidelines for identifying the fivefold-coordinated SiO$_5$ structure. Raman spectroscopy is effective for observing local structures[40,41] because it measures the eigenfrequency of a crystal as a Raman shift. Thus, we calculated the eigenfrequencies of the fivefold-coordinated SiO$_5$ structure to be about 1000 cm$^{-1}$ and 400 cm$^{-1}$. We also calculated the Raman scattering activity[42] for each vibration mode and determined that these two vibration modes can be observed by Raman spectroscopy. We next calculated the eigenfrequency of the fourfold-coordinated SiO$_4$ structure to be about 400 cm$^{-1}$. Thus, the 1000 cm$^{-1}$ eigenfrequency of a fivefold-coordinated SiO$_5$ structure is far from the eigenfrequency of the usual fourfold-coordinated SiO$_4$ structures and can be used to identify the SiO$_5$ structure. Therefore, fivefold-coordinated SiO$_5$ structures can be identified by a Raman peak at 1000 cm$^{-1}$.

In summary, we have proposed a mechanism for the storage of negative charge by potassium-ion electrets based on first-principles calculations. In this study, we modeled the manufacturing of potassium-ion electrets by first-principles MD calculations and investigated the structures and the charged states of a-SiO$_2$ with and without a potassium atom. We found that characteristic local structures, such as the fivefold-coordinated SiO$_5$ structure, appear in a-SiO$_2$ with a potassium atom. These local structures were produced by inserting a potassium atom. The potassium atom is positively charged, whereas the fivefold-coordinated SiO$_5$ structure is negatively charged. Next, we estimated the stability of the charge state after removal of the potassium atom by comparing the formation energies of each charged state as functions of the Fermi level. The negatively charged state was found to be the most stable in the experimental conditions, and the negatively charged SiO$_5$ structure is retained. Therefore, we conclude that the fivefold-coordinated SiO$_5$ structure is the physical origin of the robust storage of negative charge in potassium-ion electrets. Since the Si–O bond is extraordinarily strong, it is expected that the negative charge will be almost permanently retained. Therefore, potassium-ion electrets can be used to fabricate high-reliability vibration-powered generators. This result is in good agreement with the experimental results for a negatively charged electret. Finally, we propose that fivefold-coordinated SiO$_5$ structures can be identified experimentally because they will have a Raman peak at 1000 cm$^{-1}$.

This work was supported by the Ministry of Education, Culture, Sports, Science, and Technology, Japan, under the research project "Promoting Research on the Supercomputer Fugaku" and by the Japan Science and Technology Agency for Core Research for Evolutional Science and Technology (Grant Nos. JPMJCR15Q4 and JPMJCR19Q2).

The data that support the findings of this study are available from the corresponding author upon reasonable request.